\documentclass[11pt]{article}
\usepackage{graphicx}
\usepackage{amssymb}
\begin{document}

\title{Hidden variables or hidden theories?}

\author{A. Feoli\footnote{E-mail: feoli@unisannio.it}\\ Dipartimento di Ingegneria, Universit\`{a} del
            Sannio, \\Corso Garibaldi n. 107, Palazzo Bosco Lucarelli  \\ I-82100 - Benevento, Italy.
            \\ INFN - Sezione di Napoli - Gruppo Collegato di Salerno,
            Italy.}
\date{}
 \maketitle

 \noindent Pacs: 03.65.Ta; 03.65.Ud


\begin{abstract}
We show that a modified Relativity Principle could explain in a
"classical" way the strange correlations of entangled photons. We
propose a gedanken experiment with balls and boxes that predicts
the same distribution of probability of the Quantum Mechanics in
the case of the EPR experiment with a pair of entangled photons
meeting a pair of polarizers. In the light of this gedanken
experiment, we find an alternative description of the real EPR
experiment postulating the existence of two observers (one for
each polarizer) embedded in two locally anisotropic spacetimes. In
our model there is no need to invoke quantum non separability or
instantaneous action at distance.
\end{abstract}

\section{INTRODUCTION}

Seventy years ago Einstein, Podolsky and Rosen published a famous
paper \cite{1} in which they consider a particular two particle
state "$\psi_{12}$ that  cannot be written as a product like
$\psi_1 \psi_2$, but only as a sum of such products"\cite{ein}.
These quantum states of two particles are today  called
"entangled"  and can be described "in such a way that their global
state is perfectly defined, whereas the states of the separate
particles remain totally undefined"\cite{grangier}. Considering
pairs of entangled particles EPR showed that Quantum Mechanics
(QM) is not a complete theory and they  hoped that QM could be
improved or substituted by a theory in which the state of a
particle is better specified knowing some still hidden variables.
The probabilistic character of the quantum description of reality
would be due only to the ignorance of these hidden variables. In
1964 Bell \cite{3} showed an inequality  that is always satisfied
by local hidden variable (LHV) theories and, in some cases, is
violated by QM. The paper written by Clauser, Horne, Shimony and
Holt \cite{4} made possible a lot of experimental tests that led
to violations of their generalized Bell inequalities showing that
LHV cannot explain the experiments \cite{Bellexp}. The conclusion
is that it is impossible to construct a LHV theory that leads to
the same predictions of QM and that QM agrees with the
experiments. The hidden variable theories equivalent to QM (such
as the de Broglie - Bohm approach \cite{bohm}) must be non local.
So we must suppose, according to QM, that the entangled particles
form a "non separable" system or that, in a hidden variable
theory, an instantaneous action at distance can occur.

 In a previous paper, written with Rampone,  \cite{feo} we discussed how the noise
can alter the results of the experiments and the difficulties to
find a "fair sample" of particles to test the inequalities. In
this paper we consider the ideal case of tests without any noise
or "experimental loopholes" (locality, detection efficiency,
selection effects, etc). Although there are still some discussions
about the interpretation of the results of Aspect's like
experiments, we want to believe that quantum mechanics gives
correct predictions and so an alternative theory must reproduce
exactly the same results. We do not propose a new deterministic
theory that can complete, in the spirit of Einstein approach, the
standard Quantum Mechanics. Our aim is only to show that in a
particular case a derivation of the same predictions of quantum
mechanics is possible without any need of instantaneous action at
distance or invoking the non separability of the system of
entangled particles. We have only to suppose the existence of a
still hidden classical theory. To this aim we propose a gedanken
experiment (section 3) in which the role of the hidden theory is
played by the well known special relativity. Then we will describe
the new Relativity Principle (section 4) and the hidden theory
necessary to explain the strange correlations of quantum entangled
systems (section 5). We consider in particular two entangled
photons emitted by a source in opposite directions whose
polarizations are measured using two polarizers. The photons have
a probability to pass (or not to pass) the test of a polarizer
with some chosen polarization axis. Up to now this probability can
be predicted in the right way by the standard quantum mechanics
(section 2) and not by the local hidden variable theories.

\section{THE STANDARD THEORY}

We analyze the famous experiment of pairs of entangled photons
meeting polarizers. According with the standard interpretation of
quantum mechanics the results of the EPR - type experiments with
photons are described following these steps:
\begin{enumerate}
\item In the region A the polarizer on the left of the source,
 with a polarization axis in the position
$\theta_A$, measures the polarization of the photon N. 1.
Following the standard approach, the photons are emitted by the
source with a completely undefined polarization,  so the
probability to pass the filter is P(Yes,*) = 1/2 and of course
also P(No,*) = 1/2. If the photon passes the test, its wave
function collapses in a state with a well defined polarization
angle $\theta_A$ .

\item As the state of the second photon is entangled with the
first one, also the photon N. 2 instantaneously acquires a
polarization at the same angle $\theta_A$. According with QM and
the Malus law, it has a probability $cos^{2}(\theta_A - \theta_B)$
to pass the test of the second polarizer placed in the region B on
the right of the source, with a polarization axis in the position
$\theta_B$.

\item So, the probability that both photons pass the tests is just
 $P (Yes,Yes) = cos^{2}(\theta_A - \theta_B)/2$.

\end{enumerate}
Hence the predictions of quantum mechanics are:
$$P(*,Yes) = P(*,No) = P(Yes,*) = P(No,*) = 1/2 \eqno(1)$$
 $$P (Yes,Yes) = P(No,No)= \frac{1}{2} cos^{2}(\theta_A -
\theta_B) \eqno(2)$$
 $$P (Yes,No) = P(No,Yes) = \frac{1}{2} sin^{2}(\theta_A -
\theta_B) \eqno(3)$$ and they are confirmed by the experiments.

\section{A GEDANKEN EXPERIMENT}

We want to show a completely classical case where an observer
computes the same set of probabilities obtained in the previous
experiment with photons. In this macroscopic example we will use
balls instead of photons and boxes instead of polarizers. The
Observer A is at rest with respect to a Box with a tube inside
that is open with a large section $S_1 = \Delta x \Delta z$ on the
$xz$ plane and with a small section $S_2 = \Delta y \Delta z$ in
the $zy$ plane (Fig. 1). There is a special filter inside the tube
able to distinguish if a ball has a negative or a positive hidden
charge. On the $xz$ plane of the Box there is also a circular
instrument with an index that can be fixed at a whatever angle
$\theta_A$.
\begin{figure}
  \centering
\resizebox{\hsize}{!}{\includegraphics{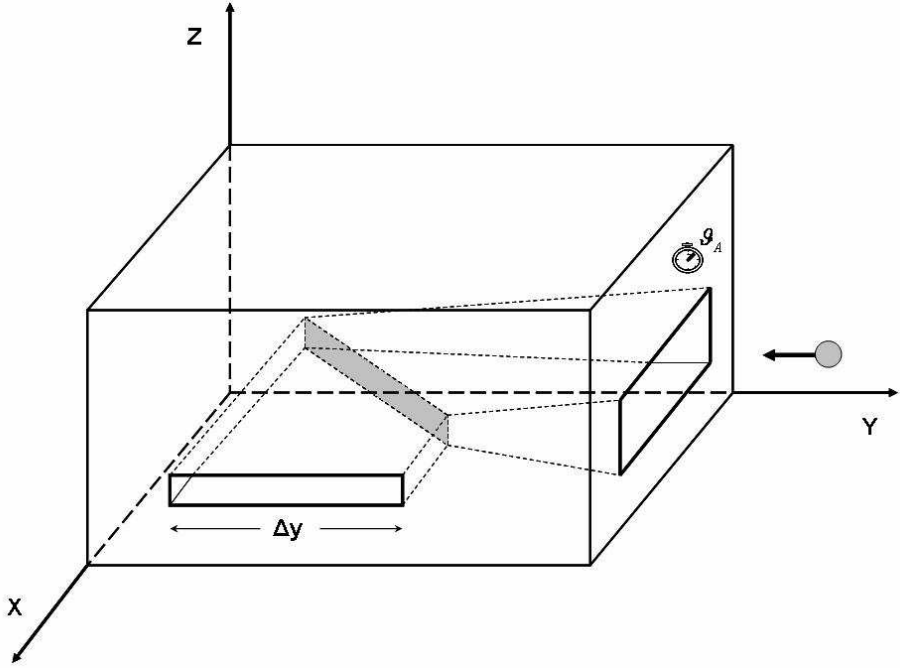}} \caption{The
ball moving towards the box of the observer A} \label{Fig. 1}
\end{figure}
The observer A sees a lot of other observers with similar boxes
that run with different constant velocity along the y axis. He
chooses his angle $\theta_A$ and one among these other observers
(that we will call B) that has the index of its circular
instrument on the position $\theta_B$. After a brief time interval
the two Boxes reach a stable  state in which  the relation between
the relative angle $\theta = \theta_A - \theta_B$ and their
relative velocity is such that $$V = c \,\, sin\theta \eqno(4)$$
where $c$ is the speed of light. Furthermore the observer A turn
on a machine that emits simultaneously two small balls that will
go into the boxes of the observer A and B respectively entering
the tubes from the $xz$ plane. The observer A must compute the
probability that the balls go out from the two tubes and,
repeating many times the experiment, must check if his prediction
is right.

The rules of the game  are the following:
\begin{enumerate}
\item The observer A knows that the machine emits with the same
frequency pairs of negatively and positively  charged balls.

\item From the instructions written on each Box, the observer A
knows the probability that a positively charged ball passes the
filter and goes out the box. It is proportional to the aperture
$\Delta y$ (Fig.1):
$$P_+(Yes) = \frac{(\Delta y)^2}{(\Delta y)^2_{MAX}} = P_-(No) \eqno(5)$$
and is also the probability that a negatively charged ball has to
be absorbed by the filter. He knows also the value of the
normalization constant $(\Delta y)_{MAX}$ (the maximal length of
the tube along the y - axis), but he ignores that all the Boxes
are constructed with this maximal length.

\item He ignores the theory of Special Relativity.

\item He can measure all the lengths, velocities and  angles.
\end{enumerate}
 In order to compute the probabilities, he measures the aperture
of his tube and he finds the value $\Delta y_A = (\Delta y)_{MAX}$
and the aperture $\Delta y_B$ of the tube of the polarizer B.
Repeating many times the experiment with different observers that
travel at different velocities with respect to him, he can argue a
phenomenological rule:
$$\Delta y_B = \Delta y_{MAX} cos\theta \eqno(6)$$ in which the
aperture of the Box B is surprisingly related to the relative
angle $\theta$ between the indexes of the two circular
instruments.

 He concludes that:
\begin{itemize}
\item $P_+(Yes,*)$ = (Prob. to find a positively charged
ball)(Prob. that it passes the filter A) $= 1/2$ and $P_+(No,*) =
0$

\item $P_-(Yes,*)$ = (Prob. to find a negatively charged ball)(Prob.
that it passes the filter A) = 0  and $P_-(No,*)=1/2$

\item $P_+(*,Yes)$ = (Prob. to find a positively charged ball)(Prob.
that it passes the filter B) = $1/2 cos^2\theta = P_-(*,No)$

\item $P_-(*,Yes)$ = (Prob. to find a negatively charged ball)(Prob.
that it passes the filter B) = $1/2 sin^2\theta = P_+(*,No)$
\end{itemize}
The probability that both balls of a pair go out from the boxes
is: $P_+(Yes,Yes)$ = (Prob. to find a pair of positively charged
balls)(Prob. that one passes the filter A) (Prob. that the other
passes the filter B). Hence:
$$P_+ (Yes,Yes) = \frac{1}{2}\left( \frac{(\Delta y_A)^2}{(\Delta y)^2_{MAX}}\right)
 \left(\frac{(\Delta y_B)^2}{(\Delta y)^2_{MAX}} \right)
 = \frac{1}{2} cos^{2}(\theta_A - \theta_B) \eqno(7)$$
 Furthermore the observer A obtains: $$P_+(Yes, No)=\frac{1}{2} sin^{2}(\theta_A -
\theta_B) \eqno(8)$$ and
$$P_+(No, Yes)= P_+(No, No)=0. \eqno(9)$$ It is very easy to compute the
corresponding probability for negatively charged balls:
 $$P_- (No,No) = \frac{1}{2} cos^{2}(\theta_A - \theta_B) \eqno(10)$$
$$P_-(No, Yes)=\frac{1}{2} sin^{2}(\theta_A - \theta_B) \eqno(11)$$ and
$$P_-(Yes, No)= P_-(Yes, Yes)=0 \eqno(12)$$
The observer B computes the probabilities in the same cases and
obtains specular results that are different from A if one
considers only positive (or negative) balls but are the same if
one considers always the sum $P_+ + P_-$. Summing the probability
of negatively and positively charged balls for each similar case,
the observer A obtains exactly the same predictions of the quantum
mechanics in the example of entangled photons (eqs.1-3).
$$P_+(*,Yes) + P_-(*,Yes) = P_+(*,No) + P_-(*,No) = 1/2 \eqno(13)$$
$$P_+(Yes,*)+  P_-(Yes,*)= P_+(No,*)+ P_-(No,*)= 1/2 \eqno(14)$$
$$P_+ (Yes,Yes)+ P_- (Yes,Yes) = P_+(No,No) + P_-(No,No)= \frac{1}{2} cos^{2}\theta
 \eqno(15)$$
 $$P_+ (Yes,No) + P_- (Yes,No) = P_+(No,Yes)+ P_-(No,Yes)= \frac{1}{2} sin^{2}\theta \eqno(16)$$
 Of course this final result will be the same predicted by the observer B
from his point of view.

 The prediction is based only on the measures
without knowing the theory of Special Relativity. From the Lorentz
contraction of lengths we know that the observer A obtains for the
aperture of the tube in the Box B the value
$$\Delta y_B = (\Delta y)_{A} \sqrt{1 - V^2/c^2} \eqno(17)$$ that
using the eq.(4) leads to the result (6). But this is only a
relativistic effect. The proper length of the aperture in B is
equal to the one in A. So the problems for the observers A and B
are due not only to hidden variables but also to a hidden theory.

\section{THE MODIFIED RELATIVITY PRINCIPLE}
The predictions of the two observers of the previous gedanken
experiment could be used to explain the results of the real
experiment with photons as an alternative to the standard QM
postulating a sort of relativity in the polarizations. So we must
think that there is an observer for each instrument. This is not a
new idea because, for example, it was applied to EPR experiments
by Smerlak and Rovelli using the words "any physical system
provides a potential observer"\cite{rovelli}. Our aim is
different. In order to save the locality, Rovelli \cite{rov},
using the formalism of QM, gives a new point of view (called
Relational Quantum Mechanics) alternative to the standard
Copenaghen interpretation. On the contrary, we want to show that
it is possible to explain in a "classical" way the EPR experiment
without using the QM formalism.

Admitting the existence of two observers A and B, an experimental
evidence is that they always obtain the same results if the
  corresponding polarizers have their optical axes with the same orientation.
 Different orientations can lead to different results of the experiments for two inertial
observers and hence for two frames of reference. So one can choose
between two alternatives:
\begin{enumerate}
\item To preserve the standard Galilean Relativity Principle compelling the
two observers to compare their results only when their
experimental devices are in the same conditions (the same
orientation for the two polarizers)

\item To conclude that all possible inertial frames of
reference are no longer physically equivalent. For each one of our
observers a preferred direction (fixed by the optical axis of the
polarizer) exists and the Relativity Principle could be changed
this way (we apply to our case a principle introduced by
Bogoslovsky \cite{Bog} in a different context):

{\it "All laws of Nature are exactly the same only in such
inertial frames of reference which have the same orientation with
respect to the preferred direction"}

\end{enumerate}
If we consider the second point of view, the observer A of our
experiment belongs to one class of equivalent frames of reference
with the preferred direction given by the unit vector
$\vec{\nu}_A$ and the observer B to another class of equivalent
frames that has in common  a different preferred direction
$\vec{\nu}_B$. The preferred direction fixes also a preferred
frame: the one with an axis (for example the y - axis) parallel to
this direction. The outcomes of the experiments can be simply
described this way:
\begin{enumerate}
\item Two quantum particles are considered "identical from the
classical point of view" (same initial conditions, etc.) only if
they are entangled.

\item If the preferred directions of the anisotropic spacetimes of
the two observers coincide $(\vec{\nu}_A =\vec{\nu}_B)$, the two
inertial reference frames are equivalent and the two observers
always obtain identical experimental results. The modified
Relativity Principle holds.

\item If $\vec{\nu}_A \neq \vec{\nu}_B$ the probability that the two
observers obtain the same results depends on the angle between the
two preferred directions and it is $P(Yes,Yes) + P(No,No) =
(\vec{\nu}_A \cdot \vec{\nu}_B)^2$.

\end{enumerate}

\section{THE RELATIVE POINT OF VIEW}
But in the light of the example of the previous section, we can
also describe the experiment with photons from a  point of view
closer to the example of balls and boxes. A source emits a pair of
entangled photons with a circular polarization. At the polarizer
the end of the electric field vector $\vec{E}$ (with a magnitude
$E$) travels around a circle and stops either when it arrives at
the direction of the optical axis or at the perpendicular
direction. As the starting direction at the source is completely
random (hidden variable), the fifty per cent of photons will
arrive at the optical axis and the fifty per cent at the
perpendicular direction. So there are two kinds of photons that
could be called positively and negatively charged photons as in
the example of balls. The probability that the photon passes is
proportional to the square of the component of its electric field
as measured by an observer along the preferred direction
$(E_y/E)^2$ of his reference frame. If the observer A is in the
preferred frame $xOy$, he has the $y$ - axis parallel to the
optical axis of the polarizer A but placed at an angle $\theta$
with respect to the optical axis of the polarizer B. He predicts
that:
$$P^A_+ (Yes,Yes) = \frac{1}{2}\left( \frac{E_y^A}{E}\right)^2
 \left(\frac{E_y^B}{E} \right)^2
 = \frac{1}{2}\cdot 1  \cdot cos^{2} \theta \eqno(18)$$
 as in the equation (7) of the section 3. Also the remaining distribution
 of probability is the same as the example with balls and boxes.
Of course an observer placed on the polarizer B with a reference
frame $x^\prime 0^\prime y^\prime$ with the $y^\prime$ - axis
parallel to the optical axis of the polarizer B computes
$$P^B_+ (Yes,Yes) = \frac{1}{2}\left( \frac{E^A_{y^\prime}}{E}\right)^2
 \left(\frac{E^B_{y^\prime}}{E} \right)^2
 = \frac{1}{2} cos^{2} \theta \cdot 1 \eqno(19)$$
and, in particular, he obtains exactly the same probability
distribution as A for the sum $P_+ + P_-$ as shown in equations
(13 -16) of the section 3.

So we must think that there is an observer and a preferred frame
for each instrument and that each observer A can assume that his
instrument allows the maximal (minimal) probability  for the
transmission of positively (negatively) charged photons. On the
other side he sees that on the other polarizer B, with a
polarization axis rotated of an angle $\theta$ (with respect to
the polarizer of the first observer) a  probability contraction
from 1 to $P^A_+ (*,Yes) = cos^2\theta$ occurs for "positively
charged" photons and a probability dilation from zero to
$P^A_-(*,Yes) = sin^2\theta$ for "negatively charged" photons.
These relations must be explained by a still hidden theory that
plays the role of special relativity of the section 3. If this
theory exists, it leads to the same probability distributions of
the previous example of balls. In the case of photons these
predictions are confirmed by the experiments so if it were
possible to distinguish between negative and positive particles we
would agree with Smerlak and Rovelli \cite{rovelli} when they
argue that in QM "different observers can give different accounts
of the same sequence of events". But if the charge of the photons
remains an hidden variable, both the observers give the same
predictions because the sum $P_+ + P_-$ for each case is an
invariant.

Probably there will be several theories that can be right to play
the role of special relativity in our example. A simple proposal
for the hidden theory could be to substitute locally the standard
special relativity with a special - relativistic theory of the
locally anisotropic spacetime where the new Relativity Principle
holds. The theory was formulated by Bogoslovsky \cite{Bog1}
\cite{Bog2}, but not applied to EPR experiments. In that framework
the length of a vector $X$ is given by (eq. n. 10 of Ref.
\cite{Bog1})
$$ || X || = \left(\frac{(\nu_i X^i)^2}{X^i X_i}\right)^{r/2}\sqrt{X^i X_i} \eqno(20)$$
where $r$  (such that $|r|<1$) is the parameter that characterizes
the magnitude of anisotropy. The interesting property of the
equation (20) is that the vector magnitude is determined not only
by its pseudo - Euclidean length, but also by its orientation with
respect to a preferred direction given by the zero vector $\nu^i=
(1, \vec{\nu})$ such that $\nu^i \nu_i =0$. In our case we must
think that the photon N. 1 is embedded in the locally anisotropic
spacetime  of the polarizer A with the preferred direction
$\vec{\nu}_A$ given by its optical axis. So the length of the
electric field vector is
$$ || E || = \left(\frac{\vec{\nu}_A \cdot \vec{E}}{E}\right)^{r}\sqrt{E^2} \eqno(21)$$
where $E^2 = E_x^2 + E_y^2 +E_z^2$. If we choose $r = 1/2$ (but it
is possible to choose another value of $r$ and then to change the
definition of the probability without changing the final result.
For example the limit case $r=1$ is very interesting.) and we
denote with $\vec{E}_A$ the electric field of the photon that
travels towards the polarizer A and with $\vec{E}_B$ the
polarization vector of the photon that travels towards the
polarizer B, we obtain the probability computed by the observer A
that both the positively charged photons pass:
$$P^A_+ (Yes,Yes) = \frac{1}{2} \left(\frac{||E_A||^2}{E^2}\right)^2 \cdot \left(\frac{||E_B||^2}{E^2}\right)^2
$$ $$ = \frac{1}{2}\left( \frac{\vec{\nu}_A \cdot
\vec{E}_A}{E}\right)^2
 \left(\frac{\vec{\nu}_A \cdot \vec{E}_B}{E} \right)^2
 = \frac{1}{2} cos^{2} \theta \eqno(22)$$
Of course the observer B is embedded in a locally anisotropic
spacetime with a preferred direction $\vec{\nu}_B$ given by the
optical axis of his polarizer B.
$$P^B_+ (Yes,Yes) = \frac{1}{2} \left(\frac{||E_A||^2}{E^2}\right)^2 \cdot \left(\frac{||E_B||^2}{E^2}\right)^2
$$ $$ = \frac{1}{2}\left( \frac{\vec{\nu}_B \cdot
\vec{E}_A}{E}\right)^2
 \left(\frac{\vec{\nu}_B \cdot \vec{E}_B}{E} \right)^2
 = \frac{1}{2} cos^{2} \theta \eqno(23)$$
and this way one can compute all the probability distribution that
will be again the same as the example of balls and boxes. So we
have shown that there is a possibility to explain the strange
behavior of entangled photons in EPR experiments, in a "classical"
way without using quantum formalism or invoking "non separability"
or instantaneous action at distance. Of course the price to pay is
to admit the existence of two observers, of a new Relativity
Principle and of a hidden theory such as the one of Bogoslovsky.

\bigskip
\noindent{\bf Acknowledgments}
\bigskip

The author is grateful to Gaetano Scarpetta for very useful
discussions and also to Davide Mele and Rossella Sanseverino.


\begin{thebibliography}{99}

\bibitem{1} A. Einstein, B. Podolsky, and N. Rosen,  Phys. Rev. 47  (1935) 777.
\bibitem{ein} A. Einstein in a Letter to Born of April 5, 1948.
\bibitem{grangier} P. Grangier,  Nature 409  (2001) 774.
\bibitem{3} J. S. Bell, Physics 1 (1964) 195.
\bibitem{4} J. F. Clauser, M. A. Horne, A. Shimony, R. A. Holt,  Phys.Rev.Lett. 23  (1969) 880.
\bibitem{Bellexp}
                 A. Aspect, P. Grangier, G. Roger,  Phys. Rev. Lett. 47  (1981)
                 460;\\
                 A. Aspect,  P. Grangier, G. Roger,  Phys. Rev. Lett. 49 (1982)
                 91;\\
                 A. Aspect, J. Dalibard, G.  Roger,  Phys. Rev. Lett. 49  (1982)
                 1804;\\
                 Z. Y. Ou, L. Mandel,  Phys. Rev. Lett. 61  (1988) 50;\\
                 J. G. Rarity,  P. R. Tapster,  Phys. Rev. Lett. 64 (1990)
                 2495;\\
                 J. Brendel, E. Mohler, W. Martienssen, Europhys. Lett. 20  (1992)
                 575;\\
                 P. G. Kwiat, A. M. Steinberg, R. Y. Chiao,  Phys. Rev. A  47 (1993)
                 2472;\\ P. R.  Tapster, J. G.  Rarity, P. C. M. Owens,  Phys. Rev.
Lett. 73  (1994) 1923;\\
                 P.G. Kwiat, K. Mattle, H. Weinfurter,  A. Zeilinger, A. V. Sergienko, Y. Shih, Phys. Rev. Lett. 75
                (1995) 4337;\\  G. Weihs, T. Jennewein, C. Simon, H.
Weinfurter, A. Zeilinger, Phys. Rev.Lett.  81 (1998) 5039;\\ M.A.
Rowe, D. Kielpinski, V. Meyer, C. A. Sackett, W. M. Itano, C.
Monroe, D. J. Wineland, Nature 409 (2001) 791.
\bibitem{bohm} L. de Broglie,  Jour. de Phys. 8 (1927) 225;\\
 D. Bohm,  Phys. Rev. 85 (1952) 166;  85 (1952) 180;\\
  L. de Broglie, {\it Nonlinear wave mechanics} (Elsevier,
Amsterdam, 1960);\\
 see also  D. Bohm, J. P. Vigier, Phys. Rev.
 96 (1954) 208; \\E. Nelson,  Phys. Rev.  150 (1966) 1079;\\
  W. Lehr, J. Park, Jour. Math. Phys. 18 (1977) 1235;\\ J. P. Vigier,
 Lett. Nuovo Cim. 24 (1979) 258 and 265;\\ J. P. Vigier,  Found. Phys. 21 (1991) 125;\\P. R.
Holland, Phys. Rep. 224 (1993) 95.
\bibitem{feo} A. Feoli, S. Rampone,  Europhys. Lett.
62 (2003) 154.
\bibitem{rovelli} M. Smerlak,  C. Rovelli, Preprint
Quant-ph/0604064 and references therein.
\bibitem{rov} C. Rovelli, Int. J. Theor. Phys. 35
(1996) 1637.
\bibitem{Bog} G. Yu. Bogoslovsky, Il Nuovo Cimento B 40
(1977) 116.
\bibitem{Bog1} G. Yu. Bogoslovsky, Il Nuovo Cimento
B 40 (1977) 99.
\bibitem{Bog2} G. Yu Bogoslovsky,  Class. Quant. Grav. 9
(1992) 569;  Phys. Part. Nucl. 24 (1993) 354;  Fortschr. Phys. 42
(1994) 143;\\ G. Yu. Bogoslovsky, H. F. Goenner, Phys. Lett. A 244
(1998) 222;  Gen. Rel. Grav. 31 (1999) 1565;  Phys. Lett. A 323
(2004) 40.
\end{thebibliography}
\end{document}